\documentclass[useAMS,usenatbib]{mn2e}

\usepackage{amsmath}
\usepackage{txfonts}
\usepackage{graphicx}
\usepackage{natbib}
\usepackage{amsfonts}
\usepackage{amssymb}
\usepackage{hyperref}
\usepackage{longtable}
\usepackage{multirow}

\bibpunct{(}{)}{;}{a}{}{,} % to follow the A&A style

\renewcommand{\hat}[1]{\oldhat{\mathbf{#1}}}

\title[Rotation periods and seismic ages of KOIs]{Rotation periods and seismic ages of KOIs - comparison with stars without detected planets from \emph{Kepler} observations.}
\author[T.~Ceillier et al.]{T.~Ceillier$^{1}$\thanks{E-mail: tugdual.ceillier@cea.fr (TC); jlvansaders@gmail.com (JvS)},
J. van Saders$^{2,3,4}$,
R.A.~Garc\'\i a$^{1,4}$, 
T.S.~Metcalfe$^{5}$,
O.~Creevey$^{6,7}$,
\newauthor
S.~Mathis$^{1}$,
S.~Mathur$^{4,5}$,
M.H.~Pinsonneault$^{4,8}$,
D.~Salabert$^{1}$ and
J.~Tayar$^{4,8}$\\
$^{1}$Laboratoire AIM, CEA/DSM -- CNRS - Univ. Paris Diderot -- IRFU/SAp, Centre de Saclay, 91191 Gif-sur-Yvette Cedex, France\\
$^{2}$Carnegie-Princeton Fellow, Carnegie Observatories, 813 Santa Barbara Street, Pasadena, California, 91101 USA\\
$^{3}$Department of Astrophysical Sciences, Princeton University, Princeton, NJ 08544, USA\\
$^{4}$Kavli Institute for Theoretical Physics, University of California, Santa Barbara, CA 93106-4030, USA\\
$^{5}$Space Science Institute, 4750 Walnut Street, Suite 205, Boulder, Colorado 80301 USA\\
$^{6}$Laboratoire Lagrange, Universit\'e de Nice Sophia-Antipolis, UMR 7293, CNRS, Observatoire de la C\^ote d'Azur, Nice, France\\
$^{7}$Institut d'Astrophysique Spatiale, Universit\'e Paris XI, UMR 8617, CNRS, Batiment 121, 91405 Orsay Cedex, France\\
$^{8}$Department of Astronomy, The Ohio State University, Columbus, Ohio 43210, USA
}
\begin{document}

\date{Accepted ---. Received ---}

\pagerange{\pageref{firstpage}--\pageref{lastpage}} \pubyear{2002}

\maketitle

\label{firstpage}

\begin{abstract}
One of the most difficult properties to derive for stars is their age. For cool main-sequence stars, gyrochronology relations can be used to infer stellar ages from measured rotation periods and HR Diagram positions.  These relations have few calibrators with known ages for old, long rotation period stars.  There is a significant sample of old \emph{Kepler} objects of interest, or KOIs, which have both measurable surface rotation periods and precise asteroseismic measurements from which ages can be accurately derived. In this work we determine the age and the rotation period of solar-like pulsating KOIs to both compare the rotation properties of stars with and without known planets and enlarge the gyrochronology calibration sample for old stars.
We use \emph{Kepler} photometric light curves to derive the stellar surface rotation periods while ages are obtained with asteroseismology using the Asteroseismic Modeling Portal in which individual mode frequencies are combined with high-resolution spectroscopic parameters. We thus determine surface rotation periods and ages for 11 planet-hosting stars, all over 2 Gyr old. We find that the planet-hosting stars exhibit a rotational behaviour that is consistent with the latest age-rotation models and similar to the rotational behaviour of stars without detected planets. We conclude that these old KOIs can be used to test and calibrate gyrochronology along with stars not known to host planets.
\end{abstract}

\begin{keywords}
stars: oscillations -- stars: rotation -- stars: evolution -- planet-star interactions
\end{keywords}

\section{Introduction}

Empirical evidence \citep{1972ApJ...171..565S} and theoretical expectation \citep{weber1967, schatzman1962} established early on that rotation and age should be related in cool main sequence stars: stars with thick convective envelopes and magnetic fields lose angular momentum in magnetized stellar winds. Because ages are among the most difficult of stellar properties to measure, while period measurements are more straightforward, a calibrated rotation period-age relationship can provide a powerful stellar diagnostic.

These ``gyrochronology'' relationships \citep{2003ApJ...586..464B, 2007ApJ...669.1167B, mamajek2008, barnes2010, 2009ApJ...695..679M, 2015Natur.517..589M} were initially calibrated using young open clusters where the stellar ages and masses are comparatively well determined. These calibrations have historically suffered from a dearth of calibrators at ages older than the Sun. 
The recent measurements of rotation periods for the open clusters NGC~6811 (1~Gyr) by \citet{2011ApJ...733L...9M} and NGC~6819 (2.5~Gyr) by \citet{2015AAS...22544909M} have made it possible to extend the validity of these relations toward intermediate ages.
However, the Sun often remains the only primary calibrating point for old field stars.
The time-domain nature of the \textit{Kepler} satellite \citep{2010Sci...327..977B} provides access to seismically inferred stellar ages and surface rotation periods for old field stars, which is a major advance for gyrochronology.
Asteroseismology is sensitive to the changes in the structure of a star throughout its evolution, and thus provides an independent measure of the stellar age. Surface rotation rates can be extracted from the modulation of the stellar light curves due to the rotation of starspots across the disk of the star, and with space data this is possible even for low-amplitude and long-period signals. Rotation periods have been extracted for large samples of \textit{Kepler} field stars \citep{2013MNRAS.432.1203M, 2013A&A...557L..10N, 2013A&A...560A...4R, 2014ApJS..211...24M}, as well as for the KOIs \citep[e.g.][]{2013ApJ...775L..11M}, and asteroseismic dwarfs \citep{2014A&A...572A..34G}.  The KOIs have been intensively studied, and as a result there are a large number of them which have precise asteroseismic measurements, among which many are quite old \citep{2015MNRAS.452.2127S}. These stars are a potentially interesting source of old calibrators for gyrochronology studies.

However, it is possible that the relationship between mass, composition, age, and rotation may be different for stars with and without detected planetary systems. The frequency and conditions under which there are star-planet interactions remains a topic of active debate. \citet{baliunas1997} reported a periodicity in Ca H $\&$ K measurements in $\tau$ Boo that matched the planetary orbital period. \citet{2009MNRAS.396.1789P} found tentative empirical evidence that tidal star-planet interactions may spin up planet hosts, although this effect is supposed to be effective only in the case of a close massive planet \citep{2012A&A...544A.124B}. Other studies have focused on the magnetic interaction of planets and host stars \citep{lanza2010, 2014A&A...565L...1P} and spin-up from planet ingestion \citep{zhang2014}. Interactions with a companion that are sufficient to alter the surface rotation period of the star will break the rotational clock and provide misleading gyrochronological ages. \citet{walk2013}, \citet{2013ApJ...775L..11M} and \citet{2015ApJ...803...69P} examined the rotation periods of KOIs, but without the benefit of precise ages they could not address this issue. The asteroseismic KOIs provide a unique window into the coevolution of stars and planets: here we can distinguish between rotation rates that are linked to the ageing process and those that have been altered by interaction.

In this paper we present rotation periods and ages for a sample of KOIs. In section \ref{obs} we describe our methods, both the extraction of the periods from the \textit{Kepler} photometry and determination of asteroseismic ages. In Section \ref{gyro} we compare the KOIs to stars that are not known to harbour planets, and show that the rotational behaviours of the two samples are similar. In Section \ref{con} we provide our conclusions.

\section{Observations and data analysis}
\label{obs}

%FIGURE 1---------------------------------------------------
\begin{figure}
\begin{center}
\includegraphics[width=9cm]{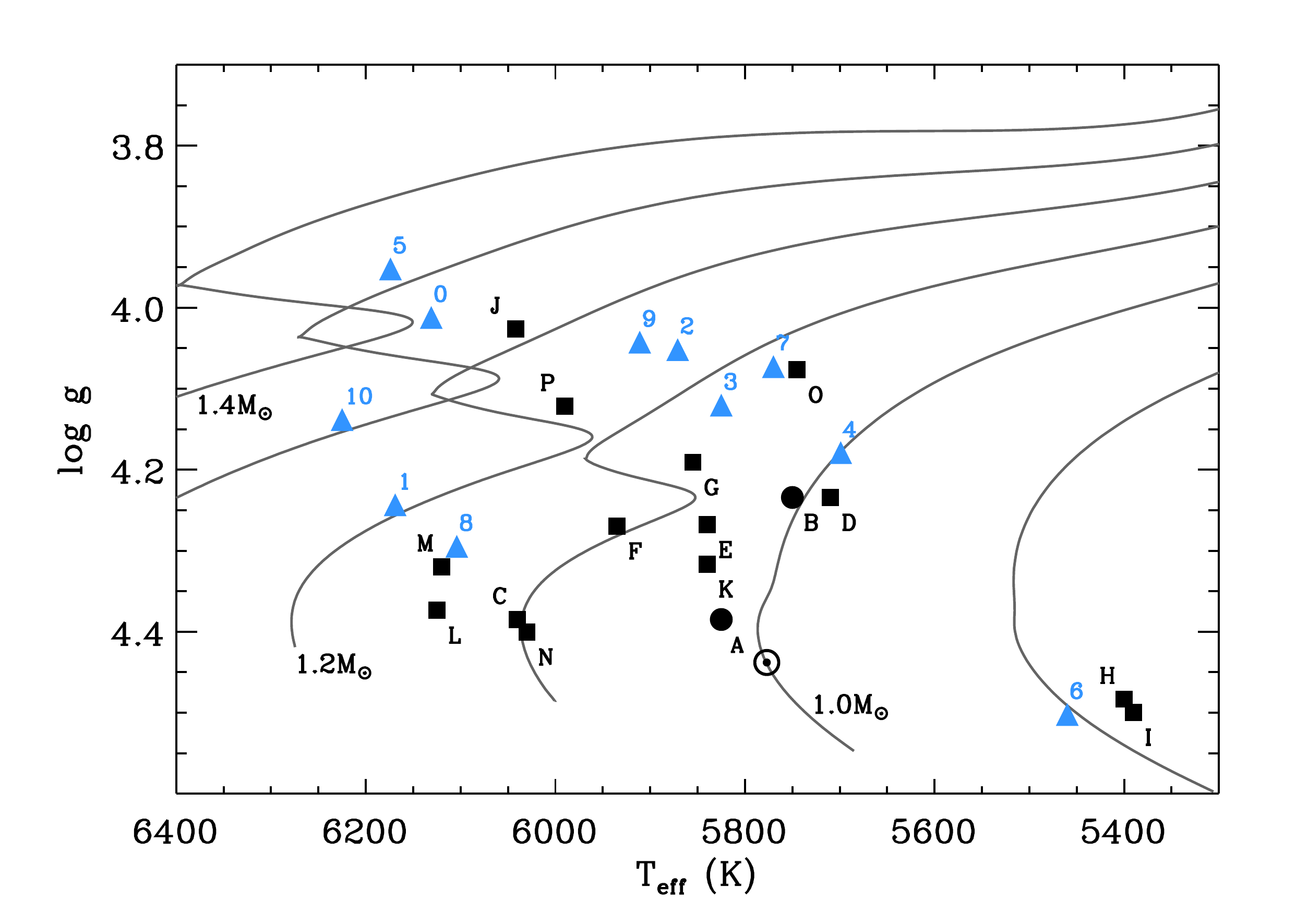}
\caption{HR diagram showing the natural logarithm of the surface gravity versus the effective temperature T$_{\rm eff}$. Squares correspond to the stars from \citet{2014A&A...572A..34G}, circles are 16Cyg A \& B from \citet{2015MNRAS.446.2959D}, and blue triangles correspond to the KOIs analysed in this work. Black lines are evolutionary tracks calculated with the ASTEC code \citep{2008Ap&SS.316..113C} for a range of masses at solar composition ($[Z/X]_\odot=0.0246$). The Sun is indicated by the $\odot$ symbol.}
\label{Fig:1}
\end{center}
\end{figure}
%FIGURE 1---------------------------------------------------

The stars we study are known planet-hosting stars from the \emph{Kepler} mission, also known as KOIs. Only 27 of these stars also have accurate asteroseismic measurements allowing for the determination of a precise age \citep[see][for more details]{2015MNRAS.452.2127S}. From the 27 KOIs of this initial sample, we are able to extract a surface rotation period for 11 stars. With the exception of KIC~9592705, which still has the status of ``candidate", all the rest of the KOIs have confirmed planets and none of them are found to be part of binary systems in the \emph{Kepler} Eclipsing Binary Catalog\footnote{\url{http://keplerebs.villanova.edu}}. The characteristics of these 11 KOIs are summarised in Table~\ref{tbl} (labelled 0 to 10). Their positions on the HR diagram are indicated as blue triangles in Fig.~\ref{Fig:1}, where the log g values has been determined asteroseismically \citep[e.g.][]{2013MNRAS.431.2419C}.

This sample of KOIs contains five single-planet systems, three 2-planet systems, and three 3-planet systems. These planets' radii range from 0.7 to 3.94 R$_\oplus$ and their semi-major axes range from 0.035 to 0.392 AU. There are therefore no hot Jupiters in our sample. All the planetary parameters can be found at the \emph{Kepler} KOI archive\footnote{\url{http://archive.stsci.edu/kepler/koi/search.php}}.

\subsection{Rotation period measurements}
\label{Sub:RotPer}

In this work we use high-precision photometry obtained by the planet-hunter \emph{Kepler} mission \citep{2010Sci...327..977B}. For the extraction of the surface rotation rate, we use \emph{Kepler} long-cadence data from quarter 0 to quarter 17 for the selected targets, cadenced at 29.4244 min \citep{2010ApJ...713L.115H}. To detect accurate rotation periods, we need light curves that are corrected from any low-frequency instrumental drift and with all quarters well-concatenated. Hence, we extracted our own aperture photometry from the so-called pixel-data files (Mathur, Bloemen, Garcia et al. in preparation) and we correct outliers, jumps, and drifts following the procedures described in \citet{2011MNRAS.414L...6G}. The smaller gaps due to missing data are then interpolated \citep{2014A&A...568A..10G,2015A&A...574A..18P}. The time series thus obtained are usually denoted as KADACS (\emph{Kepler} Asteroseismic Data Analysis and Calibration Software) light curves. These light curves are then high-pass filtered using two different triangular smoothing functions with cut-off periods of 30 and 55 days (100$\%$ of the signal is preserved for periods shorter than 30 and 55 days, and then it smoothly attenuates to zero at double periods of 60 and 110 days). The second filter, which produces noisier light curves, is dedicated to finding longer rotational periods, typically over 25 days, that can produce modulations at shorter periods in the 30d-filtered light curves. A final dataset with a longer 80 day cut-off is used to ensure that there are no longer harmonics of the signals found in the other two filters. As we are dealing with stars harboring planets, the transits are removed by folding the time series according to the orbital period of the planets and then filtering them.

The detection of rotation is achieved following \citet{2014A&A...572A..34G}. The wavelet decomposition \citep[e.g.][]{1998BAMS...79...61T,2010A&A...511A..46M} and autocorrelation function \citep[ACF, see][]{2013MNRAS.432.1203M} are calculated for each star (on both 30d and 55d filtered light curves) and a period is returned for each method. We also compute the so-called Composite Spectrum (CS, Ceillier et al., in prep.), which is the product of the ACF with the normalised GWPS (Global Wavelets Power Spectrum, see \citealt{2014A&A...572A..34G}). This CS is very sensitive  to periods appearing in both methodologies and it is a powerful diagnostic tool. For an example of these different tools, see Fig.~7 of \citet{2015MNRAS.450.3211A}. The 6 different periods returned by the different methods and datasets are then compared automatically. This methodology, evaluated using one thousand  simulated light curves, shows a high reliability with only around 5$\%$ of false positives \citep{2015MNRAS.450.3211A}. A final visual inspection is performed as the number of stars considered in the analysis is small. A single rotation period P$_{\rm rot}$ is then returned for each star in which  a clear modulation attributed to spots crossing the disk of the star is visible. Only six stars from our sample (KIC~3632418, 5866724, 6521045, 9592705, 10963065 and 11807274) have been previously analysed by \citet{2013ApJ...775L..11M}. The rotational periods they derive agree very well with the ones we provide here. A few KOI systems are close to a resonance ratio for P$_{\rm orbital}$/P$_{\rm rot}$, but we would expect to see the same or higher frequency of stars and planets with commensurate periods (within the errors) $\sim30\%$ of the time if we randomly drew planetary orbital periods from the $\sim4000$ KOIs, suggesting that these resonances are probably by chance only.

All the associated figures of this analysis can be downloaded from the CEA official website\footnote{\url{http://irfu.cea.fr/Phocea/Vie_des_labos/Ast/ast_technique.php?id_ast=3607}}.

\subsection{Asteroseismic ages}

We determine the properties of the KOI sample using the frequencies and spectroscopic properties as given in Davies et al. (submitted) as input to the Asteroseismic Modeling Portal \citep[AMP,][]{2009ApJ...699..373M,2009gcew.procE...1W} in the same configuration as described for the KASC (\emph{Kepler} Asteroseismic Science Consortium) sample in \cite{2014ApJS..214...27M}. In summary, AMP uses a parallel genetic algorithm \citep[GA,][]{2003JCoPh.185..176M} to optimize the match between stellar model output and the available set of observational constraints. The evolution models are produced with the Aarhus stellar evolution code \citep[ASTEC,][]{2008Ap&SS.316...13C}, and the oscillation frequencies are calculated with the Aarhus adiabatic pulsation code \citep[ADIPLS,][]{2008Ap&SS.316..113C}. The five adjustable model parameters include the mass ($M$), age ($t$), composition ($Z$ and $Y_{\rm i}$), and mixing-length ($\alpha$). The oscillation frequencies and other properties of each model are compared to four sets of observational constraints, including: the individual frequencies corrected for surface effects following the empirical prescription of \cite{2008ApJ...683L.175K}, the two sets of frequency ratios $r_{02}$ and $r_{010}$ defined by \cite{2003A&A...411..215R}, and the available spectroscopic constraints. A normalized $\chi^2$ is calculated for each set of constraints, and the GA attempts to minimize the mean of the four $\chi^2$ values. This allows the various asteroseismic quality metrics to be traded off against each other, while ensuring that the numerous frequencies and ratios do not overwhelm the relatively few spectroscopic constraints. 

The uncertainties are determined by using the distribution of generated models (about 80,000 for each star).  The models are ordered according to their $\chi^2$ value and the first $N$ models are used to calculate the standard deviations for $\text{T}_\text{eff}$ and Fe/H.  We adjust $N$ so that we obtain standard deviations similar to the input observational errors on these same observables.  We then use these $N$ models to define the uncertainties in the stellar parameters by calculating their standard deviations.  By using the distribution of solutions, correlations such as the one between initial helium abundance and mixing-length parameter are accounted for, in contrast to a formal uncertainty calculation where the $\chi^2$ surface can be very steep and thus can produce unrealisticly small errors that fail to account for solutions with $\chi^2$ similar to the minimum.  The obtained uncertainties are of comparable value to those found by \citet{2015MNRAS.452.2127S} with the BASTA code.

Using individual frequencies and ratios information from asteroseismic data leads to much higher precision in the determination of the stellar properties compared to those obtained using only global seismic quantities \citep[e.g.][]{2012A&A...544L..13L}. In this later case, only four or five independent observables are used to build the model \citep[e.g.][]{2014ApJS..210....1C}. In particular, the individual frequencies and ratios better constrain the internal structure of the star, and thus yield a better determination of the evolution state or its age.

\section{Gyrochronology of stars harboring planets.}
\label{gyro}

%FIGURE 2---------------------------------------------------
\begin{figure}
\begin{center}
\includegraphics[width=9cm]{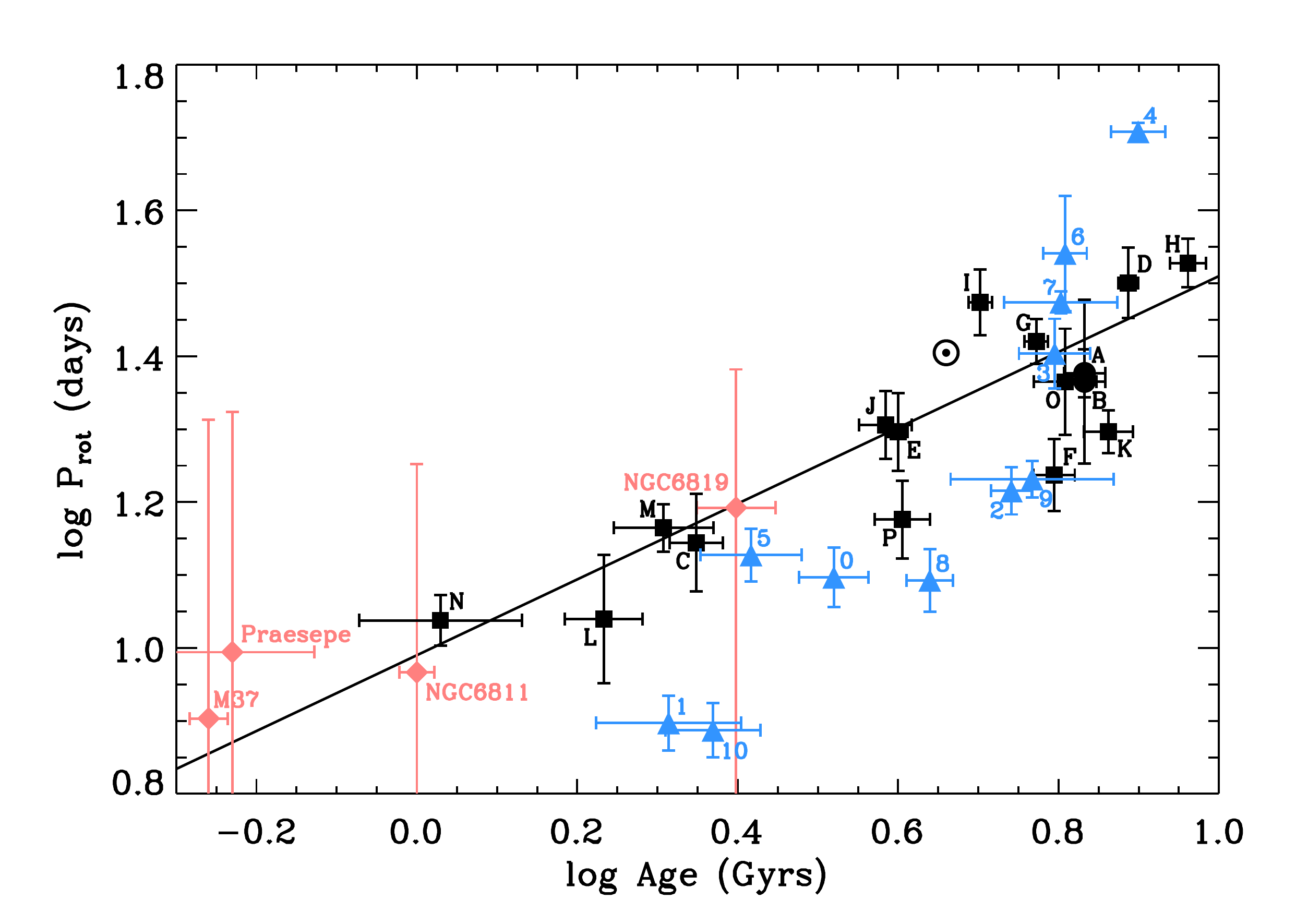}
\caption{Rotation periods, P$_{\rm{rot}}$, as a function of asteroseismic ages in a log-log space. Symbols are as in Fig.~\ref{Fig:1}. The black line is the Skumanich law fit from \citet{2014A&A...572A..34G}, using squares and circles. Magenta diamonds represent the mean values of the stars in clusters M37, Praesepe, NGC6811, and NGC6819. The vertical magenta lines correspond to the spread in rotation periods of the stars in the cluster.}
\label{Fig:2}
\end{center}
\end{figure}
%FIGURE 2---------------------------------------------------

\subsection{Rotation period-age relationships}

To evaluate the rotational behaviour of these KOIs, we compare them with a control sample of cool main-sequence dwarfs without any known planetary companions. This sample is composed of the 14 stars (black squares on Fig.~\ref{Fig:1}) from \citet{2014A&A...572A..34G} -- with precise asteroseismic ages from \citet{2014ApJS..214...27M} for 11 of the stars and \citet{2012ApJ...749..152M} for the remaining 3 --  and of 16 Cygnus A\&B (the black circles in Fig.~\ref{Fig:1}) studied by \citet{2015MNRAS.446.2959D} and which age has been derived by \citet{2012ApJ...748L..10M}, using AMP. The stars of this control sample have then ages and rotation periods derived using the same methodologies as the KOIs'. Figures of the rotation period extraction are also available at the CEA official website (see Section~\ref{Sub:RotPer}). The only exception is that the rotation period for 16Cyg~A\&B are taken from the rotational splittings of their p modes and not from the modulation of their light curve due to spots. These splittings are mostly sensitive to the outer part of the star, but also contain a small contribution from the radiative interior. For further explanations we refer to the theoretical work of \citet{2014ApJ...790..121L} and \citet{2015MNRAS.446.2959D}, and the comparisons already published between the seismically inferred surface rotation rates and the rotation period as determined from the spot modulation of the light curves \citep[e.g.][]{2013PNAS..11013267G,2015MNRAS.452.2654B,2015A&A...582A..10N}.
The derived P$_{\rm rot}$ of 16Cyg~A\&B should thus be taken as upper limits only. Most importantly, all the stars used (control sample and KOIs) have a consistent age scale, even if the absolute ages could be biased in some way. The characteristics of all the stars of the control sample are summarised in Table~\ref{tbl}. Like the KOIs, none of these stars have been found to be part of a binary or a multiple system. In the case of 16Cyg~A\&B, it is a very wide binary which does not seem to affect their rotational history \citep{2015MNRAS.446.2959D}. 

\begin{figure}
\begin{center}
\includegraphics[width=9cm]{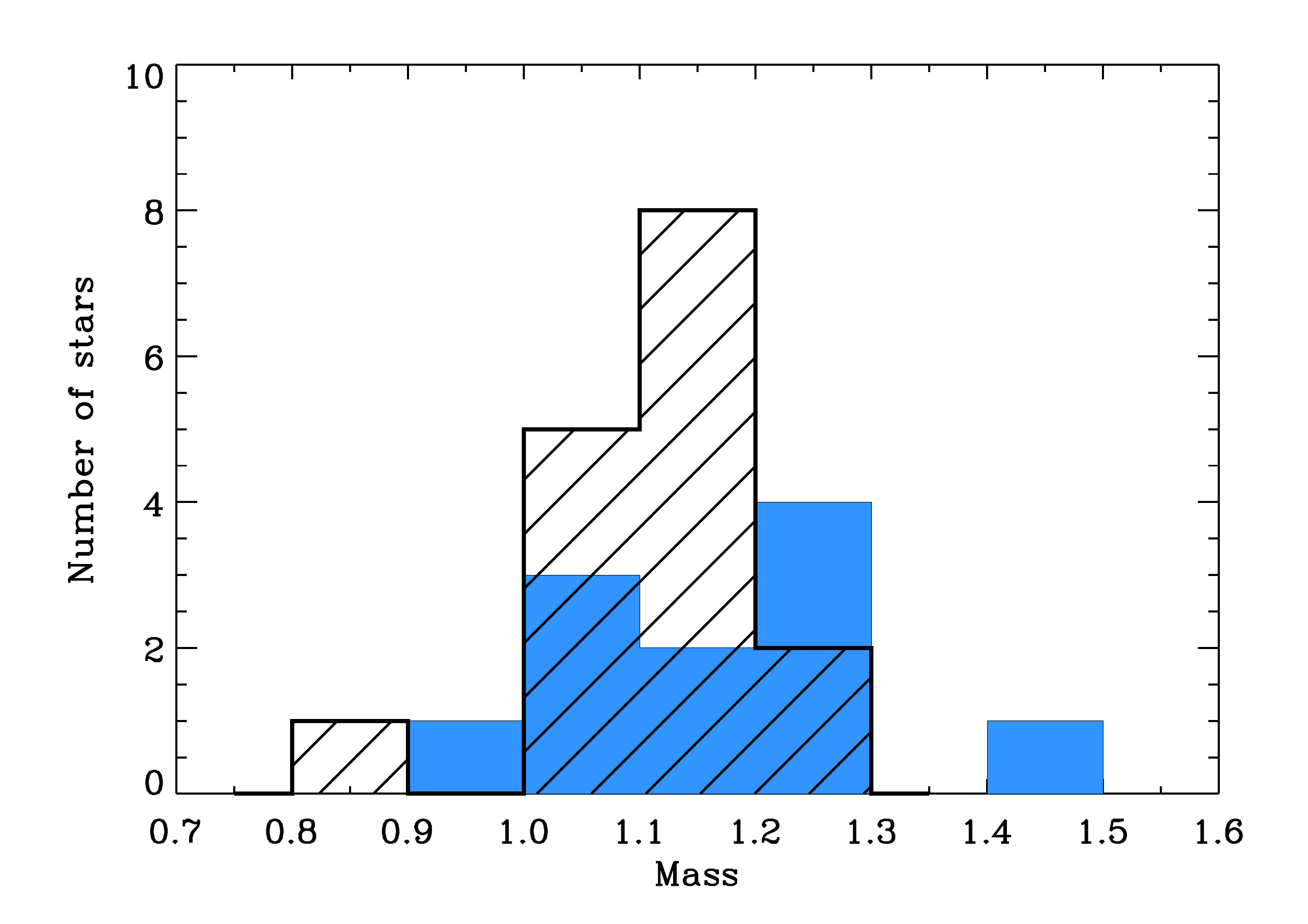}
\caption{Histograms of the asteroseismic mass for the control sample (black, striped) and the stars harboring planets (blue).}
\label{Fig:3}
\end{center}
\end{figure}

Fig.~\ref{Fig:2} presents the repartition of the KOIs and the stars of the control sample in the P$_{\rm rot}$-age space (same symbols as Fig.~\ref{Fig:1}). Due to the influence of magnetic braking and structual evolution, the rotation period becomes longer with time. There is an apparent discrepancy between the slopes and intercepts of the period-age relationships in the two samples. From this representation, one could think that KOIs have a different rotational evolution than stars without planets. But it is important to note that the fit in \citet{2014A&A...572A..34G} was performed on stars with a narrow mass range around solar, and therefore should not be applied to all the stars without considering mass and evolutionary state. More massive stars have shallower convective envelopes, so the efficiency of the magnetic braking diminishes, and we expect to observe more rapid rotation \citep[see][for more details]{2013ApJ...776...67V}. As can be seen in Fig.~\ref{Fig:3}, the KOIs (blue) have higher masses on average than the control sample.

\begin{figure}
\begin{center}
\includegraphics[width=9cm]{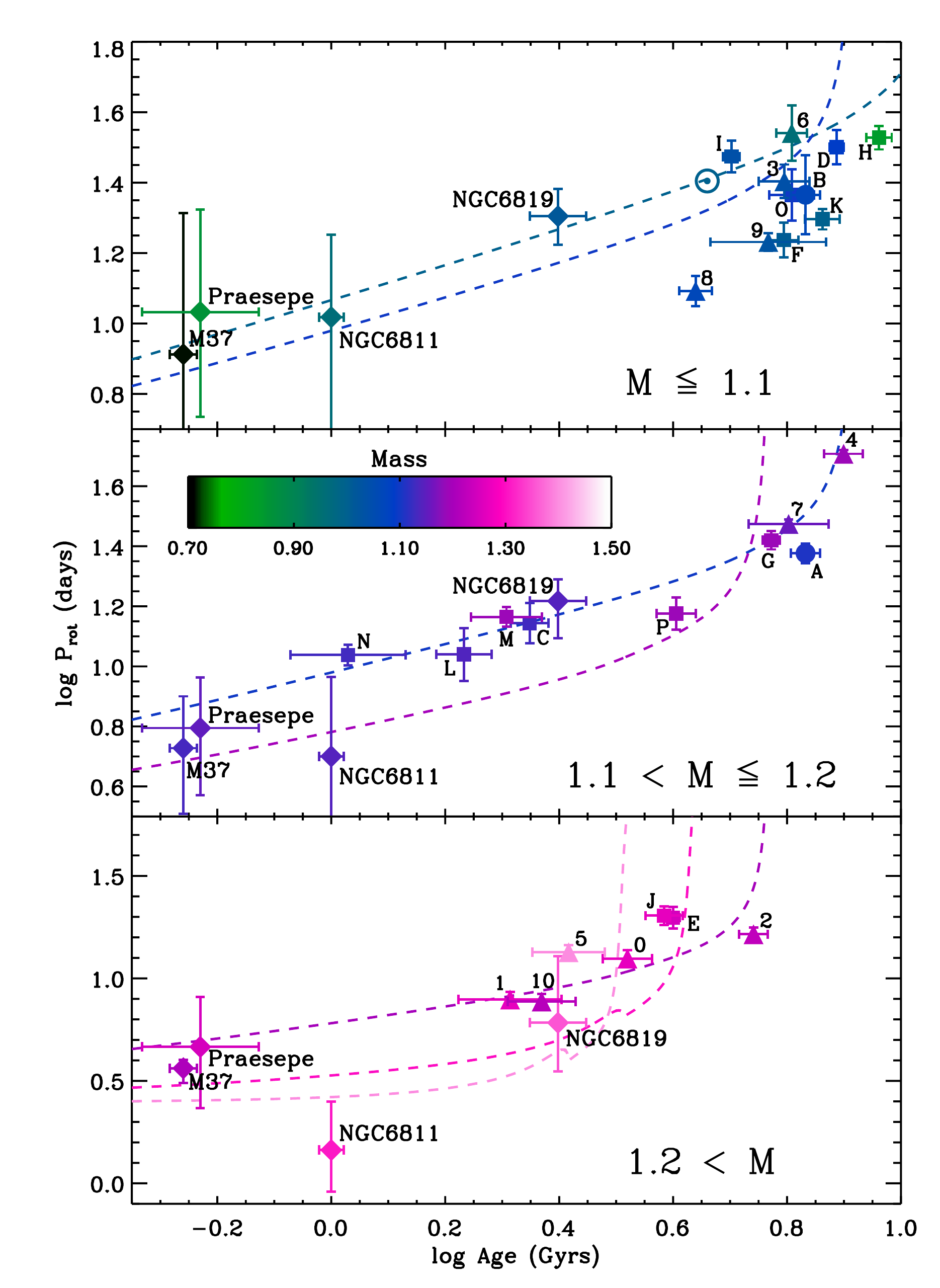}
\caption{Rotation periods, P$_{\rm{rot}}$, as a function of asteroseismic ages in a log-log space, for different mass ranges. Symbols are the same as in Fig.~\ref{Fig:1}. Colors correspond to the mass of the stars.  The colored dashed lines are evolutionary tracks from \citet{2013ApJ...776...67V}, for stars of 1.0 and 1.1~M$_\odot$ (top), 1.1 and 1.2~M$_\odot$ (middle) and 1.2, 1.3, and 1.4~M$_\odot$ (bottom). Averaged cluster values are represented by diamonds. As in Fig.~\ref{Fig:2}, the vertical lines represent the spread in rotation periods in the range of masses of each panel.}
\label{Fig:4}
\end{center}
\end{figure}

Fig.~\ref{Fig:4} divides Fig.~\ref{Fig:2} into three different panels to take into account the masses of the stars considered. The symbols are the same as Fig.~\ref{Fig:1} but the colors correspond to the mass of the stars. The stars of both samples (KOI and control) are divided into three mass ranges: $\rm{M} \leqslant 1.1 \rm{M}_\odot$, $1.1 \rm{M}_\odot < \rm{M} \leqslant 1.2 \rm{M}_\odot$ and $1.2 \rm{M}_\odot <  \rm{M}$. One can then see that all stars within the same mass range actually behave the same way. In fact, all these stars are in good agreement with the overlaid ``slow launch'' evolutionary tracks from \citet{2013ApJ...776...67V}, which set upper limits for the rotation periods.

To quantify this agreement for both samples, we compare the ratios $r=\text{P}_\text{rot, measured}/\text{P}_\text{rot, expected}$ of the KOIs and of the control sample, where $\text{P}_\text{rot, measured}$ is the measured surface rotation period and $\text{P}_\text{rot, expected}$ is the rotation period obtained from the models of \citet{2013ApJ...776...67V} for the same mass and assuming a solar metallicity and ``slow launch'' initial conditions. Using a Kolmogorov-Smirnov analysis, we find that the probability that the KOI sample's ratios and the control sample's ratios come from the same underlying distribution is $\text{p}=0.97$, which is very high. Therefore, the KOIs and the stars from the seismic control sample do not show different rotational behaviours. It is of course possible that a fraction of the stars from the control sample are actually hosting undetected planets. In this case, if these planets had an effect on the surface rotation of their host stars, we should observe a bimodality in the rotational behaviour of the control sample, which is not the case.

%FIGURE 3---------------------------------------------------
%FIGURE 3---------------------------------------------------

\subsection{Tidal analysis of KOIs}

In the context of the study of the rotational evolution of KOIs, it is also interesting to study their tidal interactions with their planetary companions. To perform this study, we follow the successful method used in \citet{2013PNAS..11013267G} and \citet{2015MNRAS.446.2959D} for the analysis of star-planet tidal interactions.

First, we consider all binary interactions between KOIs and their planet(s) ignoring planet-planet interactions in the case of multiple systems. In this framework, the key parameter to evaluate is $\alpha\equiv L_{\rm orb}/L_{*}$ \citep[see Eq.~22 in][]{1981A&A....99..126H}, which evaluates the ratio between the respective angular momentum contained in the orbit ($L_{\rm orb}$) and in the host star ($L_{*}$). As demonstrated by \citet{1980A&A....92..167H,1981A&A....99..126H}, this parameter determines the final state of a system in the case of binary interactions. If $\alpha>3$, it tends to a minimal energy state where orbits are circularized and spins are aligned and synchronized. If $\alpha<3$, planets spiral towards their host star that may be spun up because of the transfer of the orbital momentum to the star during the merger phase.  We calculate this parameter for the star-planet pairs for which a mass (or at least an upper limit) has been derived for the planet. We find that all these systems are in the second regime ($\alpha<3$). Therefore, it is important to compute the characteristic time needed for the planet to spiral onto its host star and the one for the corresponding stellar spin-up. Using Eq.~5 in \citet{2009ApJ...692L...9L} \citep[see also results obtained in][]{2009MNRAS.395.2268B} and using typical values for tidal dissipation in stars and in low-mass planets, we then obtain that only \emph{Kepler}-21b will reach the surface of its host star because of the tidal spiraling while the others will be directly engulfed by their star during the red giant phase. For all stars, the characteristic time for the potential spin-up is longer than the life-time of the stars. The caracteritics of the systems used for these calculations and the values of $\alpha$ and the spiraling times are summarised in Table~\ref{Tab:KOIs}. As a conclusion, these KOIs' rotational history should not be modified by the presence of short-period low-mass planets. This is consistent with the conclusions obtained by \citet{2012A&A...544A.124B} who demonstrated that for Sun-like stars, only extremely close gas giants orbiting highly dissipative stars impact their rotational history. As stated before, it is possible that some stars of both samples (control and KOIs) are hosting undetected planets. It is then very probable that these undetected planets are either small or far away from their host stars -- thus with a long orbital period. Consequently, their effect on the rotation of the stars should be even less important than the one of the detected planets considered here.

To conclude, we note that to have a complete picture of the dynamics/stability of multiple KOI systems, it would be necessary to take into account planet-planet interactions and the corresponding resonances (e.g. Laskar et al. 2012). It would be also important to evaluate the impact of planets on the extraction of angular momentum from the star by stellar winds \citep{2015A&A...574A..39D} and magnetic star-planet interactions \citep{2014ApJ...795...86S}. These works are out of the scope of this paper.

\section{Conclusions}
\label{con}

We study a sample of 11 pulsating  KOIs with precise asteroseismic ages ($>$~2~Gyrs) and robust surface rotation periods and compare them with a control sample of pulsating stars without known planets. From this comparison, we show that the seemingly different behaviour of the two populations can be explained by the different mass distribution of the two samples. We suggest two possible explanations for the agreement between stars with and without observed planets.

The first is that the presence of small planets does not affect the rotational evolution of the host star. 
Indeed, for our set of stars harboring planets and neglecting planet-planet interactions, the characteristic time for the potential spin-up is longer than the life-time of the stars. This result agrees well with the work of \citet{2012A&A...544A.124B} who showed that only close giant planets should influence the rotation of Sun-like stars.

Alternatively, it may be the case that nearly all stars in our samples host planets, but only a subset have detectable planets due to instrumental sensitivity or geometrical configurations. In this case, both the KOIs and control stars are drawn from the same underlying distribution, and we would not expect to see a difference in the rotational behaviour. This would be consistent with the very high planet occurrence deduced from missions like CoRoT and \emph{Kepler} \citep[][and references therein]{2015ApJ...799..180S}.

\emph{Kepler} light curves, combined with spectroscopy, can provide both surface rotation periods and asteroseismic mass, surface gravity, and age measurements for old field stars.  Our work demonstrates that KOIs and stars not known to host planets have very similar rotation-mass-age relationships. We can therefore use both classes of stars to test and extend gyrochronology relations to older field stars, which is the subject of a separate paper in preparation.

%Our work, where we combine asteroseimic ages, detection of surface rotation and gyrochronology into what we call ``Gyro\emph{seismo}chronology'', demonstrates that the gyrochronology relations hold relatively well for the cool main-sequence stars of our samples, with or without detected planets. With the recent detection of surface rotation periods for a large number of stars \citep{2013ApJ...775L..11M,2013A&A...557L..10N}, this opens the possibility to estimate the ages of a vast number of stars in a robust way. This could also have significant implications for the behaviour of subgiants and red giants and give constraints for the problematic lack of transport of angular momentum in stars \citep[e.g][]{2013ApJ...775L...1T,2013A&A...549A..75G,2013A&A...549A..74M,2013A&A...555A..54C}.

%FIGURE 4---------------------------------------------------
%FIGURE 4---------------------------------------------------

\begin{table*}
  \caption{\label{tbl} Stellar parameters. Numbers correspond to KOIs and letters to stars from the seismic control sample. In the control sample, stars indicated by a * are from \citet{2012ApJ...749..152M}, others from \citet{2014ApJS..214...27M}. 16CygA \& B are from \citet{2015MNRAS.446.2959D}.}
  \begin{tabular}{cc|*{3}{r@{$\ \pm\ $}l}|*{4}{r@{$\ \pm\ $}l}}
\multicolumn{2}{c}{} & \multicolumn{6}{c}{observed} & \multicolumn{8}{c}{inferred} \\
\hline
\# star & KIC & \multicolumn{2}{c}{$\Delta\nu$ [$\mu$Hz]} & \multicolumn{2}{c}{T$_{\rm{eff}}$ [K]} & \multicolumn{2}{c|}{P$_{\rm rot}$ [days]} & \multicolumn{2}{c}{Mass [M$_{\odot}$]} & \multicolumn{2}{c}{Radius [R$_{\odot}$]} & \multicolumn{2}{c}{$\log g$} & \multicolumn{2}{c}{Age [Gyr]} \\
\hline
0 & 3632418 &  60.86 &  0.55 &   6131 &     44 & 12.50 &  1.18 &  1.27 &  0.03 &  1.84 &  0.02 & 4.015 & 0.011  & 3.31 &  0.33  \rule{0pt}{2ex}\\
1 & 5866724 &  89.56 &  0.48 &   6169 &     50 &  7.89 &  0.68 &  1.27 &  0.06 &  1.41 &  0.02 & 4.241 & 0.007  & 2.06 &  0.43  \\
2 & 6196457 &  66.60 &  1.10 &   5871 &     94 & 16.42 &  1.22 &  1.23 &  0.04 &  1.73 &  0.02 &  4.053 & 0.011  &   5.51 &  0.32  \\
3 & 6521045 &  77.00 &  1.10 &   5825 &     75 & 25.34 &  2.78 &  1.04 &  0.02 &  1.47 &  0.01 &  4.118 &  0.016 &   6.24 &  0.64  \\
4 & 8349582 &  83.60 &  1.40 &   5699 &     74 & 51.02 &  1.45 &  1.19 &  0.04 &  1.47 &  0.05 &  4.178 &  0.003 &   7.93 &  0.62  \\
5 & 9592705 &  53.54 &  0.32 &   6174 &     92 & 13.41 &  1.11 &  1.40 &  0.05 &  2.07 &  0.02 & 3.955 &  0.019 &   2.61 &  0.38  \\
6 & 9955598 & 153.18 &  0.14 &   5460 &     75 & 34.75 &  6.31 &  0.96 &  0.03 &  0.91 &  0.01 & 4.506 & 0.003  &   6.43 &  0.40  \\
7 & 10586004 &  69.20 &  1.40 &   5770 &     83 & 29.79 &  1.02 &  1.16 &  0.05 &  1.64 &  0.02 & 4.072 & 0.010  &   6.35 &  1.03  \\
8 & 10963065 & 103.20 &  0.63 &   6104 &     74 & 12.38 &  1.22 &  1.07 &  0.03 &  1.22 &  0.01 & 4.294 &  0.004 &   4.36 &  0.29  \\
9 & 11401755 &  67.90 &  1.20 &   5911 &     66 & 17.04 &  0.98 &  1.03 &  0.07 &  1.60 &  0.04 & 4.043 &  0.022 &   5.85 &  1.37  \\
10 & 11807274 &  75.71 &  0.31 &   6225 &     75 &  7.71 &  0.66 &  1.22 &  0.05 &  1.56 &  0.02 & 4.141 & 0.009  &   2.34 &  0.32  \rule[-0.5ex]{0pt}{0pt}\\
\hline
A &16CygA &103.40 & 0.20 &   5825 &     50 & 23.80 &  1.80 &  1.11 &  0.02 &   1.24 &   0.01 & 4.295 & 0.002  &   6.80 &  0.40  
  \rule{0pt}{2ex}\\
B &16CygB &116.50 & 0.30 &   5750 &     50 & 23.20 &  6.00 &  1.07 &  0.02 &   1.13 &   0.01 & 4.362 & 0.003  &   6.80 &  0.40  \\
C &     3427720 &119.90 &  2.00 &   6040 &     84 & 13.94 &  2.15 &  1.13 &  0.04 &  1.13 &  0.01 & 4.388 & 0.003  &   2.23 &  0.17 \\
D &     3656476* & 93.30 &  1.30 &   5710 &     84 & 31.67 &  3.53 &  1.09 &  0.01 &  1.32 &  0.03 & 4.240 & 0.004 &   7.71 &  0.22  \\
E &     5184732* & 95.70 &  1.30 &   5840 &     84 & 19.79 &  2.43 &  1.25 &  0.01 &  1.36 &  0.01 & 4.270 & 0.005  &   3.98 &  0.11  \\
F &     6116048 &100.90 &  1.40 &   5935 &     84 & 17.26 &  1.96 &  1.01 &  0.03 &  1.22 &  0.01 & 4.270 & 0.007  &   6.23 &  0.37  \\
G &     7680114* & 85.10 &  1.30 &   5855 &     84 & 26.31 &  1.86 &  1.19 &  0.01 &  1.45 &  0.03 & 4.182 & 0.003  &   5.92 &  0.20  \\
H &     7871531 &153.30 &  3.60 &   5400 &     84 & 33.72 &  2.60 &  0.84 &  0.02 &  0.87 &  0.01 &4.479 & 0.007 &   9.15 &  0.47  \\
I &     8006161 &149.30 &  1.80 &   5390 &     84 & 29.79 &  3.09 &  1.04 &  0.02 &  0.95 &  0.01 & 4.502 & 0.002  &   5.04 &  0.17  \\
J &     8228742 & 61.80 &  0.60 &   6042 &     84 & 20.23 &  2.16 &  1.27 &  0.02 &  1.81 &  0.01 & 4.026 & 0.011  &   3.84 &  0.29  \\
K &     9098294 &108.80 &  1.70 &   5840 &     84 & 19.79 &  1.33 &  1.00 &  0.03 &  1.15 &  0.01 & 4.314  & 0.008  &   7.28 &  0.51  \\
L &     9139151 &117.20 &  2.10 &   6125 &     84 & 10.96 &  2.22 &  1.14 &  0.03 &  1.15 &  0.01 & 4.376 & 0.013  &   1.71 &  0.19  \\
M &    10454113 &103.80 &  1.30 &   6120 &     84 & 14.61 &  1.09 &  1.19 &  0.04 &  1.25 &  0.01 & 4.315 & 0.003  &   2.03 &  0.29  \\
N &    10644253 &123.60 &  2.70 &   6030 &     84 & 10.91 &  0.87 &  1.13 &  0.05 &  1.11 &  0.02 &4.402  &0.031   &   1.07 &  0.25  \\
O &    11244118 & 71.30 &  0.90 &   5745 &     84 & 23.17 &  3.89 &  1.10 &  0.05 &  1.59 &  0.03 &4.077  &  0.002 &   6.43 &  0.58  \\
P &    12258514 & 74.80 &  0.80 &   5990 &     84 & 15.00 &  1.84 &  1.19 &  0.03 &  1.57 &  0.01 &  4.120 & 0.001  &   4.03 &  0.32
  \rule[-0.5ex]{0pt}{0pt}\\
\hline
  \end{tabular}
\end{table*}

%TABLE ---------------------------------------------------
\begin{table*}
\begin{center}
  \caption{\label{Tab:KOIs} Parameters of the KOIs' planets and results from the tidal analysis.}
  \begin{tabular}{ccccccccc}
\hline
KIC & KOI name & \emph{Kepler} name & Planet radius & Planet mass & Period & Semi-major axis & $\alpha$ & Spiraling time\\
 &  &  & $R_\oplus$ & $M_\text{Jup}$ & days & AU & & Gyr \\
\hline
3632418 & K00975.01 & Kepler-21b & 1.59 & 0.033 & 2.79 & 0.043 & $7.6\cdot10^{-3}$ & 2.17\rule{0pt}{2ex}\\	
\multirow{3}{*}{5866724} & K00085.01 & Kepler-65c & 2.55 & 0.085 & 5.86 & 0.069 & $9.1\cdot10^{-2}$ & $>13.8$\rule{0pt}{3ex}\\	
 & K00085.02 & Kepler-65b & 1.50 & - & 2.15 & 0.035 & - & - \\	
 & K00085.03 & Kepler-65d & 1.76 & 0.0063 & 8.13 & 0.086 & $1.0\cdot10^{-2}$ & $>13.8$  \\	
\multirow{3}{*}{6196457} & K00285.01 & Kepler-92b & 3.58 & 0.19 & 13.75 & 0.119 & $4.4\cdot10^{-1}$ & $>13.8$\rule{0pt}{3ex}\\	
 & K00285.02 & Kepler-92c & 2.37 & 0.018 & 26.72 & 0.186 & $1.0\cdot10^{-1}$ & $>13.8$ \\	
 & K00285.03 & - & 1.99 & - & 49.36 & 0.28 & - & -  \\	
\multirow{3}{*}{6521045} & K00041.01 & Kepler-100c & 2.28 & 0.003 & 12.82 & 0.110 & $8.8\cdot10^{-3}$ & $>13.8$\rule{0pt}{3ex}\\	
 & K00041.02 & Kepler-100b & 1.31 & 0.023 & 6.89 & 0.073 & $3.0\cdot10^{-2}$ & $>13.8$ \\	
 & K00041.03 & Kepler-100d & 1.50 & 0.009 & 35.33 & 0.216 & $1.0\cdot10^{-1}$ & $>13.8$ \\	
8349582 & K00122.01 & Kepler-95b & 3.13 & 0.041 & 11.52 & 0.103 & $1.2\cdot10^{-1}$ & $>13.8$\rule{0pt}{3ex}\\	
9592705 & K00288.01 & - & 3.17 & - & 10.27 & 0.11 & - & - \\	
9955598 & K01925.01 & Kepler-409b & 0.98 & 0.008 & 68.96 & 0.320 & $6.5\cdot10^{-1}$ & $>13.8$ \\	
\multirow{2}{*}{10586004} & K00275.01 & Kepler-129b & 2.30 & - & 15.79 & 0.131 & - & -\rule{0pt}{3ex}\\	
 & K00275.02 & Kepler-129c & 2.40 & - & 82.20 & 0.392 & - & - \\	
10963065 & K01612.01 & Kepler-408b & 0.70 & 0.002 & 2.46 & 0.037 & $9.8\cdot10^{-4}$ & $>13.8$\rule{0pt}{3ex}\\	
\multirow{2}{*}{11401755} & K00277.01 & Kepler-36c & 3.94 & 0.025 & 16.23 & 0.128 & $8.4\cdot10^{-2}$ & $>13.8$\rule{0pt}{3ex}\\	
 & K00277.02 & Kepler-36b & 1.48 & 0.014 & 13.85 & 0.116 & $3.9\cdot10^{-2}$ & $>13.8$ \\	
\multirow{2}{*}{11807274} & K00262.01 & Kepler-50b & 1.54 & $<0.10$ & 7.81 & 0.083 & $1.3\cdot10^{-1}$ & $>13.8$\rule{0pt}{3ex}\\	
 & K00262.02 & Kepler-50c & 1.82 & $<0.11$ & 9.38 & 0.094 & $1.8\cdot10^{-1}$ & $>13.8$ \\	
\hline
  \end{tabular}
\end{center}
\end{table*}
%TABLE ---------------------------------------------------

\section*{Acknowledgments}

The authors wish to thank the entire \emph{Kepler} team, without whom these results would not be possible. Funding for this Discovery mission is provided by NASA's Science Mission Directorate. 
Authors acknowledges the KITP staff of UCSB for their hospitality during the research program Galactic Archaeology and Precision Stellar Astrophysics. MHP and JT acknowledge support from NSF grant AST-1411685.
TC, DS, and RAG received funding from the CNES GOLF and CoRoT grants at CEA. RAG also acknowledges the ANR (Agence Nationale de la Recherche, France) program IDEE (ANR-12-BS05-0008) ``Interaction Des \'Etoiles et des Exoplan\`etes''. SMathis acknowledges funding by the European Research Council through ERC grant SPIRE 647383 and the Programme National de PlanŽtologie (CNRS/INSU). SMathur acknowledges support from the NASA grant NNX12AE17G. TSM was supported by NASA grant NNX13AE91G. The research leading to these results has received funding from the European Communitys Seventh Framework Programme ([FP7/2007-2013]) under grant agreement no. 269194 (IRSES/ASK).

\bibliographystyle{mn2e}
\bibliography{BIBLIO2}

\begin{thebibliography}{60}
\expandafter\ifx\csname natexlab\endcsname\relax\def\natexlab#1{#1}\fi

\bibitem[{{Aigrain} {et~al}\mbox{.}(2015){Aigrain}, {Llama}, {Ceillier},
  {Chagas}, {Davenport}, {Garc{\'{\i}}a}, {Hay}, {Lanza}, {McQuillan}, {Mazeh},
  {de Medeiros}, {Nielsen}, \& {Reinhold}}]{2015MNRAS.450.3211A}
{Aigrain} S. {et~al.}, 2015, MNRAS, 450, 3211

\bibitem[{{Baliunas} {et~al}\mbox{.}(1997){Baliunas}, {Henry}, {Donahue},
  {Fekel}, \& {Soon}}]{baliunas1997}
{Baliunas} S.~L., {Henry} G.~W., {Donahue} R.~A., {Fekel} F.~C., {Soon} W.~H.,
  1997, ApJ, 474, L119

\bibitem[{{Barker} \& {Ogilvie}(2009)}]{2009MNRAS.395.2268B}
{Barker} A.~J., {Ogilvie} G.~I., 2009, MNRAS, 395, 2268

\bibitem[{{Barnes}(2003)}]{2003ApJ...586..464B}
{Barnes} S.~A., 2003, ApJ, 586, 464

\bibitem[{{Barnes}(2007)}]{2007ApJ...669.1167B}
{Barnes} S.~A., 2007, ApJ, 669, 1167

\bibitem[{{Barnes} \& {Kim}(2010)}]{barnes2010}
{Barnes} S.~A., {Kim} Y.-C., 2010, ApJ, 721, 675

\bibitem[{{Benomar} {et~al}\mbox{.}(2015){Benomar}, {Takata}, {Shibahashi},
  {Ceillier}, \& {Garc{\'{\i}}a}}]{2015MNRAS.452.2654B}
{Benomar} O., {Takata} M., {Shibahashi} H., {Ceillier} T., {Garc{\'{\i}}a}
  R.~A., 2015, MNRAS, 452, 2654

\bibitem[{{Bolmont} {et~al}\mbox{.}(2012){Bolmont}, {Raymond}, {Leconte}, \&
  {Matt}}]{2012A&A...544A.124B}
{Bolmont} E., {Raymond} S.~N., {Leconte} J., {Matt} S.~P., 2012, A\&A, 544,
  A124

\bibitem[{{Borucki} {et~al}\mbox{.}(2010){Borucki}, {Koch}, {Basri}, {Batalha},
  {Brown}, {Caldwell}, {Caldwell}, {Christensen-Dalsgaard}, {Cochran},
  {DeVore}, {Dunham}, {Dupree}, {Gautier}, {Geary}, {Gilliland}, {Gould},
  {Howell}, {Jenkins}, {Kondo}, {Latham}, {Marcy}, {Meibom}, {Kjeldsen},
  {Lissauer}, {Monet}, {Morrison}, {Sasselov}, {Tarter}, {Boss}, {Brownlee},
  {Owen}, {Buzasi}, {Charbonneau}, {Doyle}, {Fortney}, {Ford}, {Holman},
  {Seager}, {Steffen}, {Welsh}, {Rowe}, {Anderson}, {Buchhave}, {Ciardi},
  {Walkowicz}, {Sherry}, {Horch}, {Isaacson}, {Everett}, {Fischer}, {Torres},
  {Johnson}, {Endl}, {MacQueen}, {Bryson}, {Dotson}, {Haas}, {Kolodziejczak},
  {Van Cleve}, {Chandrasekaran}, {Twicken}, {Quintana}, {Clarke}, {Allen},
  {Li}, {Wu}, {Tenenbaum}, {Verner}, {Bruhweiler}, {Barnes}, \&
  {Prsa}}]{2010Sci...327..977B}
{Borucki} W.~J. {et~al.}, 2010, Science, 327, 977

\bibitem[{{Chaplin} {et~al}\mbox{.}(2014){Chaplin}, {Basu}, {Huber},
  {Serenelli}, {Casagrande}, {Silva Aguirre}, {Ball}, {Creevey}, {Gizon},
  {Handberg}, {Karoff}, {Lutz}, {Marques}, {Miglio}, {Stello}, {Suran},
  {Pricopi}, {Metcalfe}, {Monteiro}, {Molenda-{\.Z}akowicz}, {Appourchaux},
  {Christensen-Dalsgaard}, {Elsworth}, {Garc{\'{\i}}a}, {Houdek}, {Kjeldsen},
  {Bonanno}, {Campante}, {Corsaro}, {Gaulme}, {Hekker}, {Mathur}, {Mosser},
  {R{\'e}gulo}, \& {Salabert}}]{2014ApJS..210....1C}
{Chaplin} W.~J. {et~al.}, 2014, ApJS, 210, 1

\bibitem[{{Christensen-Dalsgaard}(2008{\natexlab{a}})}]{2008Ap&SS.316..113C}
{Christensen-Dalsgaard} J., 2008{\natexlab{a}}, Astrophys. Space. Sci., 316,
  113

\bibitem[{{Christensen-Dalsgaard}(2008{\natexlab{b}})}]{2008Ap&SS.316...13C}
{Christensen-Dalsgaard} J., 2008{\natexlab{b}}, Astrophys. Space. Sci., 316, 13

\bibitem[{{Creevey} {et~al}\mbox{.}(2013){Creevey}, {Th{\'e}venin}, {Basu},
  {Chaplin}, {Bigot}, {Elsworth}, {Huber}, {Monteiro}, \&
  {Serenelli}}]{2013MNRAS.431.2419C}
{Creevey} O.~L. {et~al.}, 2013, MNRAS, 431, 2419

\bibitem[{{Damiani} \& {Lanza}(2015)}]{2015A&A...574A..39D}
{Damiani} C., {Lanza} A.~F., 2015, A\&A, 574, A39

\bibitem[{{Davies} {et~al}\mbox{.}(2015){Davies}, {Chaplin}, {Farr},
  {Garc{\'{\i}}a}, {Lund}, {Mathis}, {Metcalfe}, {Appourchaux}, {Basu},
  {Benomar}, {Campante}, {Ceillier}, {Elsworth}, {Handberg}, {Salabert}, \&
  {Stello}}]{2015MNRAS.446.2959D}
{Davies} G.~R. {et~al.}, 2015, MNRAS, 446, 2959

\bibitem[{{Garc{\'{\i}}a} {et~al}\mbox{.}(2014{\natexlab{a}}){Garc{\'{\i}}a},
  {Ceillier}, {Salabert}, {Mathur}, {van Saders}, {Pinsonneault}, {Ballot},
  {Beck}, {Bloemen}, {Campante}, {Davies}, {do Nascimento}, {Mathis},
  {Metcalfe}, {Nielsen}, {Su{\'a}rez}, {Chaplin}, {Jim{\'e}nez}, \&
  {Karoff}}]{2014A&A...572A..34G}
{Garc{\'{\i}}a} R.~A. {et~al.}, 2014{\natexlab{a}}, A\&A, 572, A34

\bibitem[{{Garc{\'{\i}}a} {et~al}\mbox{.}(2011){Garc{\'{\i}}a}, {Hekker},
  {Stello}, {Guti{\'e}rrez-Soto}, {Handberg}, {Huber}, {Karoff},
  {Uytterhoeven}, {Appourchaux}, {Chaplin}, {Elsworth}, {Mathur}, {Ballot},
  {Christensen-Dalsgaard}, {Gilliland}, {Houdek}, {Jenkins}, {Kjeldsen},
  {McCauliff}, {Metcalfe}, {Middour}, {Molenda-Zakowicz}, {Monteiro}, {Smith},
  \& {Thompson}}]{2011MNRAS.414L...6G}
{Garc{\'{\i}}a} R.~A. {et~al.}, 2011, MNRAS, 414, L6

\bibitem[{{Garc{\'{\i}}a} {et~al}\mbox{.}(2014{\natexlab{b}}){Garc{\'{\i}}a},
  {Mathur}, {Pires}, {R{\'e}gulo}, {Bellamy}, {Pall{\'e}}, {Ballot},
  {Barcel{\'o} Forteza}, {Beck}, {Bedding}, {Ceillier}, {Roca Cort{\'e}s},
  {Salabert}, \& {Stello}}]{2014A&A...568A..10G}
{Garc{\'{\i}}a} R.~A. {et~al.}, 2014{\natexlab{b}}, A\&A, 568, A10

\bibitem[{{Gizon} {et~al}\mbox{.}(2013){Gizon}, {Ballot}, {Michel}, {Stahn},
  {Vauclair}, {Bruntt}, {Quirion}, {Benomar}, {Vauclair}, {Appourchaux},
  {Auvergne}, {Baglin}, {Barban}, {Baudin}, {Bazot}, {Campante}, {Catala},
  {Chaplin}, {Creevey}, {Deheuvels}, {Dolez}, {Elsworth}, {Garcia}, {Gaulme},
  {Mathis}, {Mathur}, {Mosser}, {Regulo}, {Roxburgh}, {Salabert}, {Samadi},
  {Sato}, {Verner}, {Hanasoge}, \& {Sreenivasan}}]{2013PNAS..11013267G}
{Gizon} L. {et~al.}, 2013, Proceedings of the National Academy of Science, 110,
  13267

\bibitem[{{Haas} {et~al}\mbox{.}(2010){Haas}, {Batalha}, {Bryson}, {Caldwell},
  {Dotson}, {Hall}, {Jenkins}, {Klaus}, {Koch}, {Kolodziejczak}, {Middour},
  {Smith}, {Sobeck}, {Stober}, {Thompson}, \& {Van
  Cleve}}]{2010ApJ...713L.115H}
{Haas} M.~R. {et~al.}, 2010, ApJ, 713, L115

\bibitem[{{Hut}(1980)}]{1980A&A....92..167H}
{Hut} P., 1980, A\&A, 92, 167

\bibitem[{{Hut}(1981)}]{1981A&A....99..126H}
{Hut} P., 1981, ApJ, 99, 126

\bibitem[{{Kjeldsen} {et~al}\mbox{.}(2008){Kjeldsen}, {Bedding}, \&
  {Christensen-Dalsgaard}}]{2008ApJ...683L.175K}
{Kjeldsen} H., {Bedding} T.~R., {Christensen-Dalsgaard} J., 2008, ApJ, 683,
  L175

\bibitem[{{Lanza}(2010)}]{lanza2010}
{Lanza} A.~F., 2010, A\&A, 512, A77

\bibitem[{{Lebreton} \& {Goupil}(2012)}]{2012A&A...544L..13L}
{Lebreton} Y., {Goupil} M.~J., 2012, A\&A, 544, L13

\bibitem[{{Levrard} {et~al}\mbox{.}(2009){Levrard}, {Winisdoerffer}, \&
  {Chabrier}}]{2009ApJ...692L...9L}
{Levrard} B., {Winisdoerffer} C., {Chabrier} G., 2009, ApJL, 692, L9

\bibitem[{{Lund} {et~al}\mbox{.}(2014){Lund}, {Miesch}, \&
  {Christensen-Dalsgaard}}]{2014ApJ...790..121L}
{Lund} M.~N., {Miesch} M.~S., {Christensen-Dalsgaard} J., 2014, ApJ, 790, 121

\bibitem[{{Mamajek} \& {Hillenbrand}(2008)}]{mamajek2008}
{Mamajek} E.~E., {Hillenbrand} L.~A., 2008, ApJ, 687, 1264

\bibitem[{{Mathur} {et~al}\mbox{.}(2010){Mathur}, {Garc{\'{\i}}a},
  {R{\'e}gulo}, {Creevey}, {Ballot}, {Salabert}, {Arentoft}, {Quirion},
  {Chaplin}, \& {Kjeldsen}}]{2010A&A...511A..46M}
{Mathur} S. {et~al.}, 2010, A\&A, 511, A46

\bibitem[{{Mathur} {et~al}\mbox{.}(2012){Mathur}, {Metcalfe}, {Woitaszek},
  {Bruntt}, {Verner}, {Christensen-Dalsgaard}, {Creevey}, {Do{\u g}an}, {Basu},
  {Karoff}, {Stello}, {Appourchaux}, {Campante}, {Chaplin}, {Garc{\'{\i}}a},
  {Bedding}, {Benomar}, {Bonanno}, {Deheuvels}, {Elsworth}, {Gaulme}, {Guzik},
  {Handberg}, {Hekker}, {Herzberg}, {Monteiro}, {Piau}, {Quirion},
  {R{\'e}gulo}, {Roth}, {Salabert}, {Serenelli}, {Thompson}, {Trampedach},
  {White}, {Ballot}, {Brand{\~a}o}, {Molenda-{\.Z}akowicz}, {Kjeldsen},
  {Twicken}, {Uddin}, \& {Wohler}}]{2012ApJ...749..152M}
{Mathur} S. {et~al.}, 2012, ApJ, 749, 152

\bibitem[{{McQuillan} {et~al}\mbox{.}(2013{\natexlab{a}}){McQuillan},
  {Aigrain}, \& {Mazeh}}]{2013MNRAS.432.1203M}
{McQuillan} A., {Aigrain} S., {Mazeh} T., 2013{\natexlab{a}}, MNRAS, 432, 1203

\bibitem[{{McQuillan} {et~al}\mbox{.}(2013{\natexlab{b}}){McQuillan}, {Mazeh},
  \& {Aigrain}}]{2013ApJ...775L..11M}
{McQuillan} A., {Mazeh} T., {Aigrain} S., 2013{\natexlab{b}}, ApJ, 775, L11

\bibitem[{{McQuillan} {et~al}\mbox{.}(2014){McQuillan}, {Mazeh}, \&
  {Aigrain}}]{2014ApJS..211...24M}
{McQuillan} A., {Mazeh} T., {Aigrain} S., 2014, ApJS, 211, 24

\bibitem[{{Meibom} {et~al}\mbox{.}(2011){Meibom}, {Barnes}, {Latham},
  {Batalha}, {Borucki}, {Koch}, {Basri}, {Walkowicz}, {Janes}, {Jenkins}, {Van
  Cleve}, {Haas}, {Bryson}, {Dupree}, {Furesz}, {Szentgyorgyi}, {Buchhave},
  {Clarke}, {Twicken}, \& {Quintana}}]{2011ApJ...733L...9M}
{Meibom} S. {et~al.}, 2011, ApJ, 733, L9

\bibitem[{{Meibom} {et~al}\mbox{.}(2015{\natexlab{a}}){Meibom}, {Barnes},
  {Platais}, {Gilliland}, {Latham}, \& {Mathieu}}]{2015Natur.517..589M}
{Meibom} S., {Barnes} S.~A., {Platais} I., {Gilliland} R.~L., {Latham} D.~W.,
  {Mathieu} R.~D., 2015{\natexlab{a}}, Nature, 517, 589

\bibitem[{{Meibom} {et~al}\mbox{.}(2015{\natexlab{b}}){Meibom}, {Barnes},
  {Platais}, {Gilliland}, {Latham}, {Mathieu}, \& {Kepler Science
  Team}}]{2015AAS...22544909M}
{Meibom} S., {Barnes} S.~A., {Platais} I., {Gilliland} R.~L., {Latham} D.~W.,
  {Mathieu} R.~D., {Kepler Science Team} K.~S.~O.~C., 2015{\natexlab{b}}, in
  American Astronomical Society Meeting Abstracts, Vol. 225, American
  Astronomical Society Meeting Abstracts, p. 449.09

\bibitem[{{Meibom} {et~al}\mbox{.}(2009){Meibom}, {Mathieu}, \&
  {Stassun}}]{2009ApJ...695..679M}
{Meibom} S., {Mathieu} R.~D., {Stassun} K.~G., 2009, ApJ, 695, 679

\bibitem[{{Metcalfe} {et~al}\mbox{.}(2012){Metcalfe}, {Chaplin}, {Appourchaux},
  {Garc{\'{\i}}a}, {Basu}, {Brand{\~a}o}, {Creevey}, {Deheuvels}, {Do{\v g}an},
  {Eggenberger}, {Karoff}, {Miglio}, {Stello}, {Y{\i}ld{\i}z}, {{\c C}elik},
  {Antia}, {Benomar}, {Howe}, {R{\'e}gulo}, {Salabert}, {Stahn}, {Bedding},
  {Davies}, {Elsworth}, {Gizon}, {Hekker}, {Mathur}, {Mosser}, {Bryson},
  {Still}, {Christensen-Dalsgaard}, {Gilliland}, {Kawaler}, {Kjeldsen},
  {Ibrahim}, {Klaus}, \& {Li}}]{2012ApJ...748L..10M}
{Metcalfe} T.~S. {et~al.}, 2012, ApJL, 748, L10

\bibitem[{{Metcalfe} \& {Charbonneau}(2003)}]{2003JCoPh.185..176M}
{Metcalfe} T.~S., {Charbonneau} P., 2003, Journal of Computational Physics,
  185, 176

\bibitem[{{Metcalfe} {et~al}\mbox{.}(2009){Metcalfe}, {Creevey}, \&
  {Christensen-Dalsgaard}}]{2009ApJ...699..373M}
{Metcalfe} T.~S., {Creevey} O.~L., {Christensen-Dalsgaard} J., 2009, ApJ, 699,
  373

\bibitem[{{Metcalfe} {et~al}\mbox{.}(2014){Metcalfe}, {Creevey}, {Do{\u g}an},
  {Mathur}, {Xu}, {Bedding}, {Chaplin}, {Christensen-Dalsgaard}, {Karoff},
  {Trampedach}, {Benomar}, {Brown}, {Buzasi}, {Campante}, {{\c C}elik},
  {Cunha}, {Davies}, {Deheuvels}, {Derekas}, {Di Mauro}, {Garc{\'{\i}}a},
  {Guzik}, {Howe}, {MacGregor}, {Mazumdar}, {Montalb{\'a}n}, {Monteiro},
  {Salabert}, {Serenelli}, {Stello}, {Steslicki}, {Suran}, {Y{\i}ld{\i}z},
  {Aksoy}, {Elsworth}, {Gruberbauer}, {Guenther}, {Lebreton}, {Molaverdikhani},
  {Pricopi}, {Simoniello}, \& {White}}]{2014ApJS..214...27M}
{Metcalfe} T.~S. {et~al.}, 2014, ApJS, 214, 27

\bibitem[{{Nielsen} {et~al}\mbox{.}(2013){Nielsen}, {Gizon}, {Schunker}, \&
  {Karoff}}]{2013A&A...557L..10N}
{Nielsen} M.~B., {Gizon} L., {Schunker} H., {Karoff} C., 2013, A\&A, 557, L10

\bibitem[{{Nielsen} {et~al}\mbox{.}(2015){Nielsen}, {Schunker}, {Gizon}, \&
  {Ball}}]{2015A&A...582A..10N}
{Nielsen} M.~B., {Schunker} H., {Gizon} L., {Ball} W.~H., 2015, A\&A, 582, A10

\bibitem[{{Paz-Chinch{\'o}n} {et~al}\mbox{.}(2015){Paz-Chinch{\'o}n},
  {Le{\~a}o}, {Bravo}, {de Freitas}, {Ferreira Lopes}, {Alves}, {Catelan},
  {Canto Martins}, \& {De Medeiros}}]{2015ApJ...803...69P}
{Paz-Chinch{\'o}n} F. {et~al.}, 2015, ApJ, 803, 69

\bibitem[{{Pires} {et~al}\mbox{.}(2015){Pires}, {Mathur}, {Garc{\'{\i}}a},
  {Ballot}, {Stello}, \& {Sato}}]{2015A&A...574A..18P}
{Pires} S., {Mathur} S., {Garc{\'{\i}}a} R.~A., {Ballot} J., {Stello} D.,
  {Sato} K., 2015, A\&A, 574, A18

\bibitem[{{Pont}(2009)}]{2009MNRAS.396.1789P}
{Pont} F., 2009, MNRAS, 396, 1789

\bibitem[{{Poppenhaeger} \& {Wolk}(2014)}]{2014A&A...565L...1P}
{Poppenhaeger} K., {Wolk} S.~J., 2014, A\&A, 565, L1

\bibitem[{{Reinhold} {et~al}\mbox{.}(2013){Reinhold}, {Reiners}, \&
  {Basri}}]{2013A&A...560A...4R}
{Reinhold} T., {Reiners} A., {Basri} G., 2013, A\&A, 560, A4

\bibitem[{{Roxburgh} \& {Vorontsov}(2003)}]{2003A&A...411..215R}
{Roxburgh} I.~W., {Vorontsov} S.~V., 2003, A\&A, 411, 215

\bibitem[{{Schatzman}(1962)}]{schatzman1962}
{Schatzman} E., 1962, Annales d'Astrophysique, 25, 18

\bibitem[{{Silburt} {et~al}\mbox{.}(2015){Silburt}, {Gaidos}, \&
  {Wu}}]{2015ApJ...799..180S}
{Silburt} A., {Gaidos} E., {Wu} Y., 2015, ApJ, 799, 180

\bibitem[{{Silva Aguirre} {et~al}\mbox{.}(2015){Silva Aguirre}, {Davies},
  {Basu}, {Christensen-Dalsgaard}, {Creevey}, {Metcalfe}, {Bedding},
  {Casagrande}, {Handberg}, {Lund}, {Nissen}, {Chaplin}, {Huber}, {Serenelli},
  {Stello}, {Van Eylen}, {Campante}, {Elsworth}, {Gilliland}, {Hekker},
  {Karoff}, {Kawaler}, {Kjeldsen}, \& {Lundkvist}}]{2015MNRAS.452.2127S}
{Silva Aguirre} V. {et~al.}, 2015, MNRAS, 452, 2127

\bibitem[{{Skumanich}(1972)}]{1972ApJ...171..565S}
{Skumanich} A., 1972, ApJ, 171, 565

\bibitem[{{Strugarek} {et~al}\mbox{.}(2014){Strugarek}, {Brun}, {Matt}, \&
  {R{\'e}ville}}]{2014ApJ...795...86S}
{Strugarek} A., {Brun} A.~S., {Matt} S.~P., {R{\'e}ville} V., 2014, ApJ, 795,
  86

\bibitem[{{Torrence} \& {Compo}(1998)}]{1998BAMS...79...61T}
{Torrence} C., {Compo} G.~P., 1998, Bulletin of the American Meteorological
  Society, 79, 61

\bibitem[{{van Saders} \& {Pinsonneault}(2013)}]{2013ApJ...776...67V}
{van Saders} J.~L., {Pinsonneault} M.~H., 2013, ApJ, 776, 67

\bibitem[{{Walkowicz} \& {Basri}(2013)}]{walk2013}
{Walkowicz} L.~M., {Basri} G.~S., 2013, MNRAS, 436, 1883

\bibitem[{{Weber} \& {Davis}(1967)}]{weber1967}
{Weber} E.~J., {Davis}, Jr. L., 1967, ApJ, 148, 217

\bibitem[{{Woitaszek} {et~al}\mbox{.}(2009){Woitaszek}, {Metcalfe}, \&
  {Shorrock}}]{2009gcew.procE...1W}
{Woitaszek} M., {Metcalfe} T., {Shorrock} I., 2009, in Proceedings of the 5th
  Grid Computing Environments Workshop, p. 1-7, p.~1

\bibitem[{{Zhang} \& {Penev}(2014)}]{zhang2014}
{Zhang} M., {Penev} K., 2014, ApJ, 787, 131

\end{thebibliography}

\bsp

\label{lastpage}

\end{document}